\documentclass[aps,prl,twocolumn,groupedaddress]{revtex4}

\newcommand{\id}{{\hbox{{\rm 1}\kern-.26em\hbox{\rm l}}}}

\begin{document}

%Title of paper
\title{Geodesics and the best measurement for distinguishing quantum states}

\author{\AA sa Ericsson}
\email[]{asae@physto.se}
\affiliation{Fysikum, Stockholms Universitet, S-106 91 Stockholm, Sweden}

\date{\today}

\begin{abstract}
  From statistical distinguishability of probability distributions, one
  can define distinguishability of quantum states. A corresponding
  measurement to perform, optimal in a definite sense, for
  distinguishing between two given states $\rho_A$ and $\rho_B$, has
  been derived by Fuchs and Caves. We show that the Bures-Uhlmann
  geodesic through the two states singles out this measurement. The
  geodesic `bounces' at the boundary of the set of quantum
  states. Whenever the geodesic hits the boundary, the state
  orthogonal to that boundary state is one of the basis states for the
  measurement.
\end{abstract}

\maketitle

Encountering two quantum states one might ask how similar they
are. Clearly one can think of ways to understand `similar' that
would make this question relevant in quantum information
processing. One reformulation of the question could be: How well can
we distinguish between the two states, with the aid of a measurement?
Orthogonal states are one-shot distinguishable---it is possible to
measure just once to know the state. But for given non-orthogonal
states there is no measurement that will discriminate between the
states with certainty. Then we can ask for the measurement that will
be the most advantageous one, in a statistical sense, for
distinguishing them. For every measurement, the two states give two
probability distributions for the outcomes. To distinguish between the
states we need to distinguish between the probability distributions.

This leads to statistical distance~\cite{Bhattacharyya,Wootters}
between states and, in a definite sense, a best distinguishing
measurement. Fuchs and Caves~\cite{F&C} have derived an expression for
this distance, and also for the operator corresponding to the optimal
measurement. The distance turned out to be the same as the geodesic
distance of the Bures-Uhlmann
metric~\cite{Bures,Uhlmann1,Uhlmann2,Uhlmann3} on the set of quantum
states. In this letter we first introduce these results and some facts
about the Bures-Uhlmann metric.  This will lead the way to our
result. We will show how the optimal measurement is determined by the
Bures-Uhlmann geodesic connecting the two quantum states: the geodesic
`bounces' at the boundary of the set of quantum states, at states
orthogonal to the basis states of the measurement operator.

Every measurement that can be performed on a quantum system can be
described by a POVM---a positive operator valued measure. This is a
complete set of non-negative Hermitian operators $E_i$ acting on an
$N$-dimensional Hilbert space $\mathcal{H}_N$; $i$ indexes the
measurement outcomes and completeness means $\sum_i E_i = \id$ (the
identity operator). If the POVM-elements are orthogonal
one-dimensional projectors, $E_i=|e_i\rangle\langle e_i|$, we have a
von Neumann measurement, with a corresponding observable
$O=\sum_{i=1}^N \lambda_i E_i$. Upon measuring, the outcome $i$ will
occur with probability $p_i=\textrm{Tr} E_i\rho$, where $\rho$ is the
density operator describing the state of the system. Thus, two states
$\rho_A$ and $\rho_B$ will produce two probability distributions
$p^{(A)}$ and~$p^{(B)}$.

Sampling from these distributions a finite number of times will give
frequencies that differ somewhat from the probabilities. From the size
of the statistical fluctuations, a measure of distinguishability
between probability distributions can be found. This is the
statistical distance $d(p^{(A)},p^{(B)})$, by
Bhattacharyya~\cite{Bhattacharyya} and Wootters~\cite{Wootters}, given
by
\begin{equation}
  \cos d(p^{(A)},p^{(B)}) = \sum_i \sqrt{p_i^{(A)} p_i^{(B)}} \ .
\end{equation}
The corresponding Riemannian metric is known as the Fisher-Rao metric.

Different POVMs give different probability distributions, and thus
different statistical distance. Take the measurement that gives the
maximal statistical distance. Then, by definition, that distance is
the distance $d(\rho_A,\rho_B)$ between quantum states:
\begin{equation}
  d(\rho_A,\rho_B) \equiv \max_{\{E_i\}} \ \arccos 
  \Big( \sum_{i} \sqrt{\textrm{Tr}E_i\rho_A}
                      \sqrt{\textrm{Tr}E_i\rho_B} \Big) \ .
\end{equation}
Fuchs and Caves~\cite{F&C} showed that the maximization yields
\begin{equation} \label{distance}
  d(\rho_A,\rho_B) = \arccos \textrm{Tr} \sqrt{\rho_A^{1/2}\rho_B\;\rho_A^{1/2}} \ ,
\end{equation}
where $\sqrt{\rho}=\rho^{1/2}$ is the unique non-negative operator
such that $\rho^{1/2}\rho^{1/2}=\rho$.  This expression looks
asymmetric, but one can show that it is not.  The most separated
states are orthogonal states, which rest at the distance $\pi/2$ from
each other. For pure states $|\psi_A\rangle$ and $|\psi_B\rangle$ we
have $d=\arccos |\langle \psi_A|\psi_B \rangle|$, the angle in Hilbert
space between the two vectors.

It may be noted that the same trace expression appears in the
fidelity~\cite{Jozsa}:
\begin{equation}
  F(\rho_A,\rho_B)=
  \left(\textrm{Tr}\sqrt{\rho_A^{1/2}\rho_B\;\rho_A^{1/2}}\,\right)^2 \ .
\end{equation}
It is a choice for a quantity that should measure the accuracy of
transmission of a signal via a quantum channel.

What is more of our concern here is that the distance
$d(\rho_A,\rho_B)$~(\ref{distance}) is the same as the geodesic
distance in the set of density operators, according to the
Bures-Uhlmann metric. Shortly we will see how this metric is
constructed. But first we will take a look at the best
measurement---the measurement found to correspond to the maximal
distance~(\ref{distance})~\cite{F&C}. It is a measurement of the
observable
\begin{equation} \label{M}
  M = \rho_A^{-1/2} \sqrt{\rho_A^{1/2}\rho_B\;\rho_A^{1/2}} \; \rho_A^{-1/2} \ .
\end{equation}
This optimal measurement is unique, except for some special
cases. (These are when $M$ has degenerate eigenvalues, or when the
states are non-invertible, i.e. when they are boundary states.)
Although we have an explicit expression for the best distinguishing
measurement, it is not easy to determine what it is for given density
operators. Operator square roots are rather unwieldy to
compute. However, this operator has appeared earlier. It is a part of
the prescription for Bures-Uhlmann geodesics. We will take advantage
of this connection to get a new description of the best measurement.

It should also be mentioned that formula~(\ref{M}) is an operator
mean. Operator means, fulfilling a set of reasonable criteria, can be
defined for two positive operators $A$ and $B$~\cite{Ando}. The
geometric mean,
\begin{equation}
  A \# B = A^{1/2} \sqrt{A_{}^{-1/2} B\, A^{-1/2}} \; A^{1/2} \ ,
\end{equation}
is one of these. Thus, we see that the optimal observable~(\ref{M}) is
the geometric mean of $\rho_A^{-1}$ and $\rho_B$:
\begin{equation}
  M =  \rho_A^{-1} \# \, \rho_B \ .
\end{equation}

Now we will turn to the Bures-Uhlmann
metric~\cite{Bures,Uhlmann1,Uhlmann2,Uhlmann3}. It is obtained from a
(kind of) fibre bundle construction in the Hilbert-Schmidt space. From
this space of operators $W$, acting on $\mathcal{H}_N$, we have the
projection
\begin{equation}
  W \longrightarrow \rho = W W^\dagger \ ,
\end{equation}
to the base manifold of positive operators. ($W^\dagger$ denotes the
Hermitian conjugate of $W$.) The fibres are obtained by right
multiplication of the unitary group, since $W$ and $WU$, for unitary
$U$, will be projected to the same operator. In the bundle space we
define distances $D(W_A,W_B)$ by
\begin{equation}
  D^2(W_A,W_B) = \textrm{Tr}(W_A-W_B)(W_A^\dagger-W_B^\dagger)
\end{equation}
--a Euclidean distance. In the base manifold of positive operators, the
distance between two operators is defined as the length of the
shortest path between the corresponding fibres in the bundle
space. Here we are only interested in the set of normalized density
operators, which means projections of operators $W$ on the unit
sphere, $\textrm{Tr}WW^\dagger=1$, in the Hilbert-Schmidt space. With
this restriction, the geodesic distance we get in the set of density
operators is given by equation~(\ref{distance}).

The physical interpretation of this construction is that of
state-purification. Every mixed state $\rho$ can be purified in a
larger Hilbert space; the system is regarded as s subsystem of a
bipartite system. The Hilbert-Schmidt space takes the role of the
larger state space. Every operator $W$ represents a pure state vector,
and the reduced density operator for the subsystem is $W
W^\dagger$. $W$ is said to be a purification of $\rho$, if $\rho=W
W^\dagger$. Consequently the whole fibre $WU$ consists of
purifications of $\rho$. The distance between the states $\rho_A$ and
$\rho_B$ of the subsystem should not be larger than between any two
purifications $W_A$ and $W_B$. This is assured by finding the shortest
path between the fibres of purifications~\cite{Uhlmann3}.

Geodesics on the unit sphere (i.e. great circles) in the
Hilbert-Schmidt space can be expressed in the following way:
\begin{equation} \label{Wgeodesic}
  W(t) = W_0 \cos t + \dot{W}_0 \sin t
  \ , \quad  0 \leq t < 2\pi \ ,
\end{equation}
with normalization and orthogonality conditions,
$\textrm{Tr}W_0W^\dagger_0=\textrm{Tr}\dot{W}_0\dot{W}^\dagger_0=1$
and $\textrm{Tr}(W_0\dot{W}^\dagger_0+W^\dagger_0\dot{W}_0)=0$.  For
this curve to project to a geodesic in the set of density operators,
it is also required that it is everywhere perpendicular to the
fibres---the `horizontality condition'. This reads
\begin{equation} \label{horizontality}
  \dot{W}^\dagger_0 W_0 = W^\dagger_0 \dot{W}_0 \ .
\end{equation}
The geodesic of density operators is then the curve
\begin{equation} \label{geodesic}
  \rho(t) = W(t) W^\dagger(t)
  \ , \quad  0 \leq t < \pi \ .
\end{equation}
The projected curve will do two turns, since $W$ and~$-W$---always
resting on the same geodesic---will be projected to the same density
operator. Hence, the range for $t$ is halved.

Let us consider the geodesic between the states $\rho_A$
and~$\rho_B$. Assume $W_A$ to be a preimage of $\rho_A$. Using the
`horizontality condition'~(\ref{horizontality}) (and positivity of
$\rho_A$), it can be shown that the operator $W_B$, that is, the
preimage of~$\rho_B$, should be given by
\begin{equation} \label{A_B}
  W_B=M W_A \ ,
\end{equation}
where $M$ is the positive operator given by~(\ref{M})---the operator
corresponding to the best distinguishing measurement.

Now we start the geodesic at $\rho_A$ (assumed to be invertible) and
let it go through $\rho_B$, resting at the geodesic distance $d$
away. This means that we set
\begin{equation}
\left\{ \begin{array}{l}
  W_A = W(0) \ , \\
  W_B = W(d) \ , \ \textrm{where} \ 
  \cos d = \textrm{Tr}\sqrt{\rho_A^{1/2}\rho_B\;\rho_A^{1/2}} \ .
\end{array} \right.
\end{equation}
Inserting in~(\ref{Wgeodesic}), we can solve for the geodesic $W(t)$
in terms of $W_A$ and $W_B$:
\begin{equation}
   W(t) = W_A \cos t + (W_B - W_A \cos d)\frac{\sin t}{\sin d} \ .
\end{equation}
Alternatively, if we use $W_B=M W_A$, we can express $W(t)$ in
terms of $W_A$ and $M$:
\begin{equation} \label{W(t)}
   W(t) = \left(\id \cos t + (M - \id \cos d)\frac{\sin t}{\sin d}\right)W_A \ .
\end{equation}
Here $d$ should be understood as given by $\cos d=\textrm{Tr}
M\rho_A$. Finally we have, for the projected curve~\cite{Sommers},
\begin{equation} \label{Xgeodesic}
\begin{array}{c}
   \rho(t) = X(t) \rho_A X(t) \ , \\[6pt]
   \textrm{where} \quad
   X(t) = \left(\id \cos t + (M - \id \cos d)\frac{\sin t}{\sin d}\right) \ .
\end{array}
\end{equation}
In this form we can think of a geodesic as given by a starting point
$\rho_A$ and a positive matrix $M$, determining the direction from
$\rho_A$.

This shows that there is a close relation between the geodesic and the
operator $M$ for the optimal distinguishing measurement. It is then
natural to ask: To what extent does the geodesic determine the best
measurement?  First we note that the measurement is given by the
eigenbasis of $M$, while the eigenvalues are superfluous
information. To answer the question we need to know more about the
geodesics. In one of Uhlmann's papers~\cite{Uhlmann3} it is explained
how the Bures-Uhlmann geodesics 'bounces' at the boundary of the set
of density operators. We will now investigate this feature with the
purpose of proving a geometric description of the optimal measurement.

At the boundary of the set of quantum states the matrices $\rho$ have
at least one zero eigenvalue, hence $\det\rho=0$. Consider the states
on a geodesic, given by equation (\ref{Xgeodesic}).
\begin{equation}
\begin{array}{c}
  \det \rho(t) = 0
  \quad \Leftrightarrow \quad
  \det X(t) = 0 \\[6pt]
  \Leftrightarrow \quad
  \det \left( M - \id x \right) = 0 \ , \\[6pt]
  \textrm{where} \quad  x = \cos d - \cot t \sin d
\end{array}\end{equation}
The solutions for $x$ of this equation are the eigenvalues $\lambda_i$
of $M$.  For the corresponding values of $t$, the states $\rho(t)$ lie
on the boundary. These states are
\begin{equation} \label{rhoboundary}
  \rho(t_i) =
  \left( M - \id \lambda_i \right) \rho_A \left( M - \id\lambda_i \right)
  \frac{\sin^2 t_i}{\sin^2 d} \ .
\end{equation}
($t_i$ is given by the equation $\lambda_i=\cos d - \cot t_i \sin d$.)
There are $N$ or less boundary states, since it is the same as the
number of different eigenvalues of $M$.  And the number of zero
eigenvalues of $\rho(t_i)$ is the same as the degeneracy of the
relevant eigenvalue of $M$. When the projected curve reaches the
boundary of the set of density matrices it will bounce back into the
interior. After $N$, or sometimes less, bounces, the curve will return
to its starting point. Thus, the geodesic will consist of a number of
segments with endpoints at the boundary of the set of density
matrices.

From the expression~(\ref{rhoboundary}), it is easily recognized what
states are orthogonal to geodesics boundary
states~$\rho(t_i)$---they are nothing but the eigenvectors
$|m_i\rangle$ of $M$:
\begin{equation}
  \langle m_i|\rho(t_i)|m_i\rangle = 0 \ .
\end{equation}
If all eigenvalues $\lambda_i$ are distinct, we get $N$ boundary
states, each with one zero eigenvalue, which singles out the $N$ basis
states $|m_i\rangle$. If there are degeneracies we get less boundary
states, but the sum of the zero eigenvalues is still $N$. And a state
$\rho(t_i)$ with $n$ zero eigenvalues is orthogonal to an
$n$-dimensional subspace of the pure states. In this subspace any
basis can be chosen; $M$ is diagonal in anyone of them. Thus, the
geodesic singles out the basis states---the optimal measurement is
fully determined by the geodesic.

For two-level systems this provides a practical method for finding the
optimal measurement. Every pair of states then lies in a disc in the
Bloch ball, which is isometric to a round
hemisphere~\cite{Hubner}. The geodesics on the hemisphere are just
great circles and the relation to the disk is by orthographic
projection. In this case the geodesic's endpoints at the `equator'
are the basis states of the measurement. For higher dimensional
Hilbert spaces, the picture is much more complex.

In conclusion we have seen how the Bures-Uhlmann geodesics in the set
of quantum states bounces at the boundary in a set of $N$, or less,
states. If there are $N$ boundary states, the states orthogonal to
these form an orthonormal basis in the Hilbert space. If there are
less than $N$ boundary states there are just enough lower rank states,
so that it is again possible to form an orthonormal basis of states
orthogonal to the boundary states. We have shown that such a basis is
the basis of an optimal distinguishing measurement: the best
measurement---in a specific statistical sense---to perform for
distinguishing between two quantum states lying on a segment of the
geodesic. This result is a conceptually interesting characterization
of the optimal distinguishing measurement.

\begin{acknowledgments}
  I gratefully acknowledge many discussions with Ingemar Bengtsson. He
  has taught me about the Bures-Uhlmann geometry and encouraged me to do
  this work. 
\end{acknowledgments}

\vspace{4pt} \emph{Note added in proof:} An expression for the
geodesics, similar to equation~(\ref{Xgeodesic}), has also been
derived by Barnum~\cite{Barnum}.

\end{document}